\documentclass[useAMS,usenatbib]{mn2e}
\usepackage{graphicx}

\title[Spectral synthesis analysis and radial velocity study of the
northern F-, G-, and K-type flare-stars]
{Spectral synthesis analysis and radial velocity study of the northern
  F-, G-, and K-type flare-stars\thanks{Partly based on
   observations collected at the German-Spanish Astronomical Center,
   Calar Alto, operated jointly by Max-Planck Institut f\"ur
   Astronomie and Instituto de Astrof\'isica de Andalucia (CSIC), and
   partly based on observations taken with the 2-m-Alfred Jensch
   telescope of the Th\"uringer Landessternwarte Tautenburg.}} 
\author[B. K\"onig and E.W. Guenther and M. Esposito and
  A. Hatzes]{B. K\"onig$^{1}$\thanks{E-mail: bkoenig@phyast.pitt.edu;
    guenther@tls-tautenburg.de (TLS); massi@tls-tautenburg.de(TLS); 
    artie@tls-tautenburg.de(TLS)} and E. W. Guenther$^{2}$ and
  M. Esposito$^{2}$ and A. Hatzes$^{2}$\\ 
$^{1}$University of Pittsburgh, Dept. for Physics and
              Astronomy, Allen Hall 406, 3941 O'Hara St, Pittsburgh,
              PA 15260, USA\\
 $^{2}$Th\"uringer Landessternwarte Tautenburg, 
              Sternwarte 5, D-07778 Tautenburg, Germany}
\begin{document}

\date{}

\pagerange{\pageref{firstpage}--\pageref{lastpage}} \pubyear{2005}

\maketitle

\label{firstpage}

\begin{abstract}
In this publication we present a study of the general physical,
chemical properties and radial velocity monitoring of young active
stars. We derive temperatures, $\log g$, $[Fe/H]$, $v \sin i$, and
$R_{\rm spec}$ values for eight stars.  The detailed analysis reveals
that the stars are not as homogeneous in their premier physical
parameters as well as in the age distribution. In 4/5 we found a
periodic radial velocity signal which origins in surface features the
fifth is surprisingly inactive and shows little variation.
\end{abstract}


\section{Introduction}
Ten years ago the first extra solar planet was discovered around
51~Peg \citep{mayor1995} and today we know of more than 160 planets
discovered with radial velocity (RV) measurements. Previous RV planet
search programs have concentrated on old and inactive stars. It is now
generally believed that the orbits of planetary systems change in
time. This is of crucial importance for the formation of planets and
the evolution of extra solar planetary systems
\citep[e.g.][]{weidenschilling1996, alibert2005}. Therefore it is
necessary to study the orbits of young extrasolar planets and compare
them to old ones. We therefore monitor the RV of 40 young stars at the
Th\"uringer Landessternwarte (TLS) and 80 young stars with HARPS at
3.6\,m telescope at ESO/La Silla.

Young stars are active and one source for active stars within 100\,pc
is the catalogue of flare stars published by \citet{gershberg}. In this
paper we discuss the G- and K-stars in this catalogue of the northern
hemisphere. We present the results of the spectral synthesis analysis
performed to derive the stellar parameters, to compare these to the
results achieved by other groups, and to put the stars into a general
picture, e.g. identify the stellar association the stars origin
from. We present here the results of the RV measurement of five
stars. The remaining three were too faint and too far south from the
TLS to measure a sufficient amount of RV data points.

The publication is organised as follows: A short description of the
analysing methods is followed by the results of this analysis for the
individual stars, followed by an overview of the RV monitoring and the
findings, followed by a discussion and conclusions.

\section{Data reduction and spectral synthesis analysis}
We observed the northern flare star sample 2000-2001 from Calar Alto
using the high resolution \'echelle spectrograph FOCES
\citep{Pfeiffer1998} mounted on the 2.2\,m telescope. Data reduction
and analysis was carried out using the reduction pipeline written in
IDL especially for this fibre coupled spectrograph.

For the spectral synthesis analysis we used the model atmospheres
MAFAGS. For a detailed description of the methods see
\citet{Fuhrmann1997}. \citet{grupp2004} has extended the program but has
also given a detailed description of the state of the art of the model
atmosphere code MAFAGS. A number of eight stars of the northern
Gershberg sample were analysed in total. As described in
\citet{Fuhrmann1997} we deduce the effective temperature from the
Balmer line wings, the surface gravity from the iron ionisation
equilibrium and the wings of the Mg~Ib lines. The analysis is
performed strictly relative to the Sun.

\subsection{Analysis of individual stars}
In this paper we want to treat each star individually because, as we
will see later, some stars have very individual features. We have
summarized our analysis in Tab.~\ref{tab:parameters}.{\sl In addition
to our analysis we have listed $\log(L_{\rm X}/L_{\rm bol})$ in
Tab.~\ref{tab:parameters} for each star which is a measure of the
stellar activity (see e.g. \cite{stelzer2005}). The X-ray luminosity
was calculated from the ROSAT PSPC count rates listed in the ROSAT
bright and faint source catalogs (Voges et al. 1999, 2000)} and
converting it to a X-ray luminosity using an energy conversion factor
of $\log({\rm ecf})=11.05\pm0.05$. For each object we will also
discuss the results of other authors using narrow-band photometry or
spectral synthesis analysis.

\begin{table*}
\caption{Spectral parameters of the flare-stars derived by spectral synthesis
  analysis. $\log(L_{\rm X}/L_{\rm bol})$ is a measure for the activity of the stars.}
\begin{tabular}{lccrrcrrccc}
\hline
name       & $T_{\rm eff}$ & Mass      & $\log g$      & $[Fe/H]$ & $M_V$ & $M_{bol}$ & $v \sin i$     & $R_{\rm spec}$ & $P_{\rm rot}$ & $\log(L_{\rm X}/L_{\rm bol})$ \\
           & [K]           & M$_\odot$ &               &                 & [mag] & [mag]     & [km\,s$^{-1}$]         & [R$_\odot$]    & [days]   &     \\
\hline                                                                                                                                                                          
HIP 1803   & $5740\pm70$   & 1.0       & $4.34\pm0.10$ & $+0.14\pm0.07$  & 4.59 &  4.71      & $6.39\pm1.00$  & $1.03\pm0.03$ & & $-4.49\pm0.11$ \\ 
HIP 18512A & $4600\pm100$  & 0.7       & $4.60\pm0.20$ & $-0.03\pm0.15$  & 8.06 &  6.53      & $2.10\pm2.00$  & $0.69\pm0.03$ & &  $-3.82\pm0.12$ \\
HIP 26779  & $5260\pm70$   & 0.9       & $4.50\pm0.10$ & $+0.09\pm0.07$  & 6.12 &  5.43      & $3.10\pm1.00$  & $0.88\pm0.02$ & $\sim\,14$ &  $-4.53\pm0.11$ \\
EK Dra     & $5700\pm70$   & 0.9       & $4.37\pm0.10$ & $-0.16\pm0.07$  & 7.60 &  4.79      & $16.50\pm1.00$ & $1.00\pm0.02$ & $2.767\pm0.005$  &  $-3.48\pm0.09$ \\ 
HIP 73555  & $4990\pm70$   & 3.4       & $2.40\pm0.10$ & $+0.01\pm0.07$  & 3.51 & -0.96      & $4.10\pm1.00$  & $18.5\pm0.5$ & $200\pm15$  &  $-6.56\pm0.21$ \\
HIP 98698  & $4730\pm70$   & 0.8       & $4.60\pm0.10$ & $-0.04\pm0.07$  & 7.47 &  6.39      & $2.50\pm1.00$  & $0.70\pm0.02$ & & $-4.82\pm0.20$ \\
HN Peg     & $5950\pm70$   & 1.0       & $4.35\pm0.10$ & $-0.07\pm0.07$  & 5.94 &  4.51      & $9.90\pm1.00$  & $1.05\pm0.02$ &  multiple & $-4.32\pm0.11$ \\ 
HIP 114385 & $5830\pm70$   & 1.1       & $4.38\pm0.10$ & $+0.08\pm0.07$  & 7.11 &  4.65      & $6.10\pm1.00$  & $1.02\pm0.02$ & $2.95\pm0.05$  & $-3.37\pm0.13$ \\  
\hline
\end{tabular}
\label{tab:parameters}
\end{table*} 

\subsubsection{HIP~1803,  [GKL99] 15, HD 1835, BE Cet}
HIP~1803 is a well known G-type star at a distance of
$20.39\pm0.38$\,pc and as often referred to is a young solar twin. The
age estimates are $\sim 600$\,Myrs which are confirmed by the fact that
it is a Hyades super-cluster member \citep{Montes2001}.  From the
space motion \citet{Montes2001} concludes that HIP~1803 is a member of
the Hyades Super Cluster. \citep{Gaidos2000} give kinematic
measurements of $(U/V/W)_{\rm LSR} =
(-26.7\pm0.7/-3.3\pm0.3/7.1\pm0.1)$\,km\,s$^{-1}$ and
\citet{mishenina2004} of $(U/V/W)_\odot =
(-35.7/-14.7/-0.2)$\,km\,s$^{-1}$ with respect to the Sun ($U$ is
defined to be positive in the Galactic anti-centric direction, and the
adopted solar motion with respect to the LSR is
$(U,V,W)=(-10.2,+5.2,+7.2)$\,km\,s$^{-1}$).

The star has a distant companion at a separation of 3860\,AU and
$\Delta V = 6.1$\,mag with the name BD~-13~60B. \citet{Favata1997}
find a lithium equivalent width of $W({\rm Li}) = 76$\,m\AA, and an
abundance of $\log N({\rm Li}) = 2.46$, at a temperature of $T_{\rm
eff} = 5732$\,K. They consider the star kinematically young. Longterm
photometric monitoring of HIP~1803 from \citet{Cutispoto1995} and the
$v \sin i$ from \citet{Pasquini1991} lead to a radius of $1.06
R_\odot$ at $i$ close to $90^\circ$.

The star was also target for the search of a debris disc
\citep{greaves2004} but no disc could be detected at 850\,$\mu$m with
a flux limit of $2\sigma$ (2\,mJy) corresponding to a total of $\le
0.02$\,M$_\odot$ of dust per star.

The FOCES spectra were observed on Dec 01 2001 with a signal-to-noise
ratio of 250. We measure a lithium EW of 72.5\,m\AA\ which transforms
to a lithium abundance of $\log N({\rm Li})= 2.30$, and the core of
H$\alpha$ is filled-in to a level of 5\%. Also the cores of calcium
H\&K-lines are filled. All in all the star is quite active. The
magnetically sensitive iron line at 6173\,\AA\ is asymmetric which
lead to the conclusion that there is a strong magnetic field on the
surface. Our spectral synthesis leads to the stellar parameters listed
in Tab.~\ref{tab:parameters}. The difference between the predicted
spectroscopic distance and the parallax measurement of {\sl Hipparcos}
\citep{ESA97} is only 2.8\%.

\citet{Gaidos2000} have measured a rotation period of
$7.81\pm0.01$\,days. Using our estimate of the radius and the $v \sin
i=6.39$\,km\,s$^{-1}$ we estimate an inclination of $0.054^\circ$ and
$v = 6675$\,m\,s$^{-1}$.

\begin{table*}
\caption{Spectral parameters of HIP~1803 derived using the spectral
  synthesis analysis methods described before (first row of the
  table). The last ten rows summarise the work of the other authors,
  where most of the data was compiled by Cayrel de Strobel et al. (1997)
  from several sources. Malagnini et al. (2000) has listed the same stars in
  their work. The spectral parameters were used to calculate the
  spectroscopic distance and compare to the {\sl Hipparcos} distance to
  investigate the accuracy of the analysis.}
\begin{tabular}{lllllccc}
\hline
 $T_{\rm eff}$   & $\log g$ & [Fe/H]    & m$_V$ & $M_{\rm bol}$ & $d_{\rm HIP}$ & $d_{\rm spec}$ & Reference \\
$[{\rm K}]$ &               & [dex]         & [mag] & [mag]        & [pc]          & [pc]        & \\
\hline
$5740\pm70$ &  $4.37\pm0.10$  & $0.13\pm0.07$ & 6.39  & 4.65       & 20.39         &  20.95      & this work \\
\hline
5771        &  4.44         & 0.15          &       & 4.81         &               &  19.5       & Katz \\
$5675\pm60$ & $4.31\pm0.12$ & $0.17\pm0.05$ &       & 4.56         &               &  21.7       & G    \\ 
5860        & 4.4           & $-0.09$       &       & 4.65         &               &  21.1       & M, C \\
5793        & 4.5           & 0.19          &       & 4.95         &               &  18.4       & M, C \\
5860        & 4.4           & 0.28          &       & 5.65         &               &  21.2       & M, C \\
5793        & 4.6           & 0.24          &       & 5.20         &               &  16.4       & M, C \\
5793        & 4.5           & 0.2           &       & 4.95         &               &  18.4       & M, C \\
5673        & 4.22          & $-0.01$       &       & 4.34         &               &  24.0       & M, C \\
5781        & 4.40          &               &       &              &               &             & Glu  \\
5800        & 4.60          & 0.24          &       & 4.69         &               &  20.7       & A \\
\hline
\end{tabular}

Katz: \citet{Katz}, G: \citet{Gaidos2002}, M: \citet{Malagnini}, C: \citet{CayreldeStrobel1997} and references therein, Glu: \cite{Glushneva2000}, A: \cite{Abia1988}
\label{tab:hip1803}
\end{table*} 

\subsubsection{HIP 18512, [GKL99] 104, HD 24916 A}
Relying on kinematic data, \citet{Madsen2002} and \citet{Montes2001}
list HIP~18512 as a member of the Ursa Major
association. They estimate a parallax $\pi = 63.3 \pm 2.0$\,mas and
measured by {\sl Hipparcos} of $\pi = 63.41 \pm 2.0$\,mas, respectively.
\citet{Favata1997} estimated a temperature of $T_{\rm eff} = 4394$\,K
and they measured lithium equivalent width of $W({\rm Li}) = 4$\,m\AA\
and an abundance of $\log N({\rm Li}) = -0.55$. They consider the star
kinematically young.

The results of our analysis of the FOCES spectra observed Dec 01 2001
and with a SNR of 300 give the parameters: $T_{\rm eff} \sim
4600\pm100$, $[Fe/H] = 0.03\pm0.07$, $\log g = 4.60\pm0.10$. We do not
detect a trace of the lithium absorption line at 6707\,\AA.

\subsubsection{HIP 26779, [GKL99] 145, HD 37394}
The space motion derived by \citep{Gaidos1998} show that the star is a
member of the local association. The stellar parameters they derive
are $T_{\rm eff} = 5200$\,K, $W({\rm Li}) = 1.3\pm3.2$\,m\AA\ which
lead to a lithium abundance of $\log N({\rm Li}) = 0.03 \pm 0.03$, $v
\sin i = 4.0$\,km\,s$^{-1}$ and an inclination range $i= 45^\circ -
90^\circ$.  Analysing the FOCES spectra, we get a very good agreement
of the spectroscopic distance with the {\sl Hipparcos} distance of
only 0.5\%.

\subsubsection{EK Dra, HIP 71631, HD 129333, [GKL99] 306}
EK~Dra is an interesting young binary where it is possible to solve the
Keplerian orbit and derive the true masses \citet{koenig2005}. The masses of
the two stars are $0.90\pm0.10$\,M$_\odot$ and $0.50\pm0.10$\,M$_\odot$,
respectively.

We also performed a spectral synthesis analysis in course of that
publication. The results are summarised in Tab.~\ref{tab:parameters}.
As mentioned in \citet{koenig2005}, EK~Dra has a large
overabundance.  The measured lithium equivalent width in the observed
spectrum is $190\pm10$\,m\AA. Modelling the lithium line this leads to a lithium
contents of $\log N({\rm Li})= 3.30\pm0.05$.

\subsubsection{HIP 73555, [GKL99] 312,  HD 133208}
The recent classification in the SIMBAD database list it as G8IIIa which would
mean that it is a post-main-sequence subgiant. Since the star was listed in
the catalogue it was observed with high resolution spectroscopy using the FOCES
instrument at Calar Alto in Feb. 13th 2001.

The spectral synthesis analysis reveal the nature of the star as a
subgiant given the estimated $\log g$ using the Fe~I/Fe~II and Fe/Mg
ionisation equilibrium.

\begin{figure}
\includegraphics[width=0.35\textwidth, angle=90]{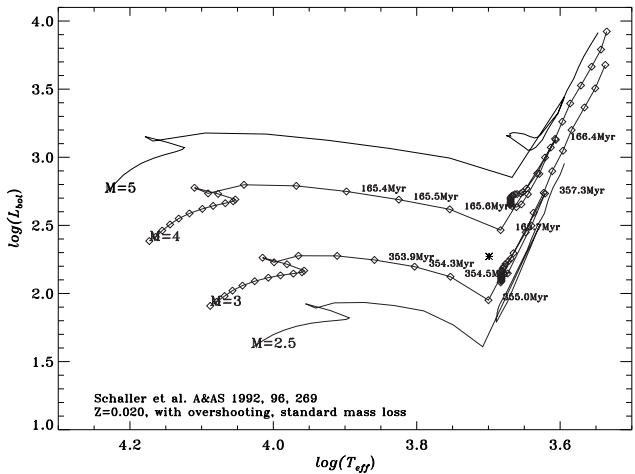}
\caption{Schaller et al. (1992) post-main-sequence tracks and age-marks. 
    Indicated with a $*$ is the position of HIP~73555.}
\label{fig:schaller}
\end{figure}

Placing the star into an Herzsprung-Russell diagram (see
Fig.~\ref{fig:schaller}) using post-main-sequence tracks of
\citet{Schaller1992} we estimate the mass of $3.4\pm0.2\,{\rm M}_\odot$
and an age of 240\,Myrs. This would imply that it is still a young
star, but because of its higher mass the evolution time scales are
smaller and yet it has reached the post-main-sequence.
Calculating a spectroscopic parallax and comparing it with the
{\sl Hipparcos} parallax the difference is 3.6\%. The kinematics of this
star reveal a space velocity of $(U/V/W)=(6.6/-17.1/-1.2)$\,km\,s$^{-1}$. This
is consistent with the Pleiades, $\alpha$~Per, IC~2602, and Coma
Berenice moving groups, which are summarised as the local association.
Since the local association is a conglomerate of comoving groups with
different ages, the age estimated from the post-main-sequence tracks
of \citet{Schaller1992} cannot be confirmed or disproved.

For the subgiant HIP~73555 radius measurement are available and listed
in \citet{Richichi2002} and references therein. By interferometry they
measure a diameter of $2.99 \pm 0.01$\,mas with a parallax of $\pi =
14.91\pm0.57$\,mas this leads to a diameter of $D = 43.0\pm2.0\,{\rm
R}_\odot$, or radius $R=21.5\pm1.0\,{\rm R}_\odot$. Our measured
temperature can be used to predict a radius: $18.5\pm0.5\,{\rm
R}_\odot$. The radius predictions are in agreement, so we are
confident that our measured effective temperature is reasonable.

\begin{table*}
\caption{Spectral parameters of HIP~73555 derived using the
  spectral synthesis analysis methods described before (first row of
  the table). The last three rows summarise the work of the other
  authors. The spectral parameters were used to calculate the
  spectroscopic distance and compare to the {\sl Hipparcos} distance to
  investigate the accuracy of the analysis. } 
\begin{tabular}{lllllcccll}
\hline
Reference & T$_{\rm eff}$   & log g         & [Fe/H]    & M$_V$ & $M_{\rm bol}$ & $d_{\rm HIP}$ & $d_{\rm spec}$ &  $R_{\rm measured}$ & $R(T_{\rm eff})$ \\
          & $[{\rm K}]$ &               & [dex]         & [mag] & [mag]         & [pc]          & [pc]           & [R$_\odot$]        &  [R$_\odot$] \\
\hline
this work & $4990\pm70$ & $2.40\pm0.10$ & $0.01\pm0.07$ & 3.51  & $-0.96$       & 67.7          & 69.5           & $21.5\pm1.0$       & $18.5\pm0.5$ \\
\hline
B         & 5000        &  2.00         & $-0.07$       &       & $-1.04$       &               & 111            & & 19.1\\
McW       & 5150        &  3.06         & $-0.13\pm0.10$&       & 0.47          &               & 36             & & 9.0\\
B\&G      & 4929        &  2.30         & 0.04          &       & $-1.24$       &               & 75             & & 21.6\\
\hline
\end{tabular}

K: this work, B: \citet{Brown1989}, McW: \citet{McWilliam1990}, B\&G: \citet{Bell1989}

\label{tab:hip73555}
\end{table*} 

\subsubsection{HIP 98698, HD~19007, [GKL99] 381}
HIP~98698 was observed May 24 2002 from Calar Alto with FOCES.
The spectrum of the star shows a young main-sequence star with a temperature
of $4720\pm80$\,K. The surface gravity ($\log g =4.60 \pm 0.10$) and the iron
abundance is $[Fe/H]=-0.04\pm0.07$.

\subsubsection{HN Peg, HIP 107350, HD 206860, [GKL99] 410}
HN~Peg was observed from Calar Alto using the FOCES spectrograph in
Nov. 29 2001. What guides the eye in this spectrum is the strong
lithium absorption feature, the filling-in of H$\alpha$ of about 10\%
and the line-broadening due to the high projected rotational velocity
$v \sin i$. The activity level namely the line-filling would support
an age of 100-300\,Myrs. \citet{Chen2001} studied the lithium
abundance: $EW({\rm Li}) = 109.5$\,m\AA\ which transforms to an
abundance of $\log N({\rm Li}) = 2.73$ derived by non-LTE
calculations. The space velocity is $(U/V/W) =
(5.1/-16.4/3.1)$\,km\,s$^{-1}$.  The stars is a member of the local
association.

The spectral syntheses analysis reveals a star close to the
main-sequence on the zero-age main-sequence (ZAMS). The measured
lithium equivalent width in the observed spectrum is
$100.8\pm0.8$\,m\AA. Modelling the lithium line this leads to a
lithium contents of $\log N({\rm Li})= +2.81\pm0.09$. Thus, like
EK~Dra, HN~Peg also has a large overabundance of lithium. The stellar
parameters are listed in Tab.~\ref{tab:parameters}.

\subsubsection{HIP~114385, HD 218739 and HIP~114379, [GKL99] 440, HD 218738}
The stars HIP~114385 and HIP~114379~A\&B form a triple system where
HIP~114385 and HIP~114379 from a visible binary system and
HIP~114379~A\&B is a spectroscopic and adaptiv optics (AO)
binary. HIP~114385 and HIP~114379 are located at the same distance at
$34.06\pm2.31$\,mas and $39.56\pm7.67$\,mas respectively. They are
comoving. Note the big error in the parallax measurement of {\sl
Hipparcos}. For HIP~114379 this clearly shows that the companion
disturbed the precision of the parallax measurement. But also the
stray light from HIP~114379 ($\pi = 39.56 \pm 7.67$\,mas) disturbed
the parallax measurement of HIP~114385 ($\pi = 34.06 \pm 2.31$\,mas)
which can be noticed by the relative big error.

For HIP~114379 the following data is available from the catalogue of
nearby star metallicities \citep{Zakhozhah1996} obtained from
photometric UBV data ($\pi = 28.7$\,mas, $[Fe/H] = -0.30$). This
result must be regarded with care because the star is a binary and the
data has not been corrected for this.

For the visual companion (HIP~114385) no data from that
catalogue is available. We measure the metallicity of the companion star
to be $[Fe/H] = 0.08\pm0.07$, much higher than its companion. It is
not clear, why the companion should have a different metallicity. We
can only guess that the metallicity determined by photometry is not
correct.

Its lithium is already burned which makes the star older than the
Pleiades, but its activity indicate an age similar to the Ursa Major
cluster.

\subsection{Distance determination}
\begin{figure}
\includegraphics[width=0.48\textwidth, angle=0]{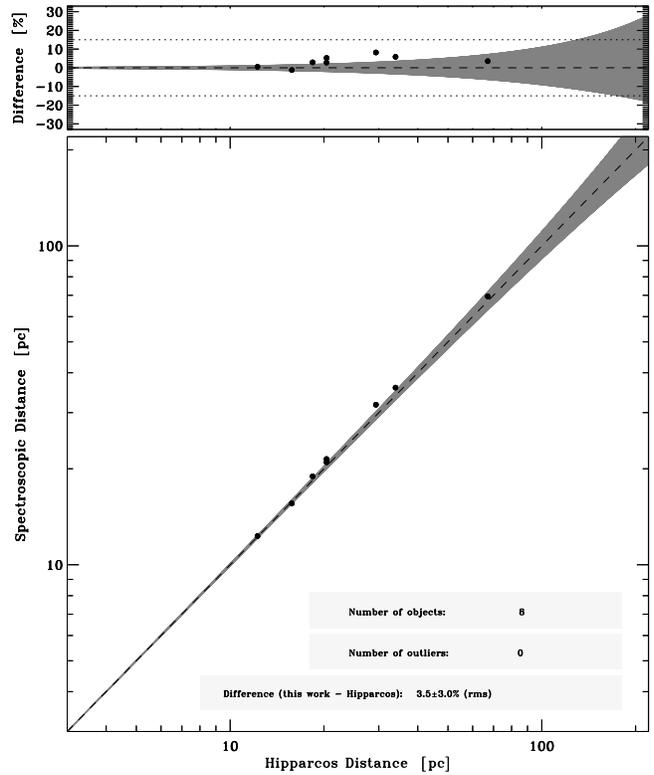}
\caption{{\sl Hipparcos} distance versus spectroscopic distance of the
flare-stars. Underlayed in gray is the average errorbar of the {\sl Hipparcos}
parallax measurements. The lines marking the 15\% level correspond to a $2
\sigma$ confidence level.}
\label{fgk_acc}
\end{figure}

Using the stellar parameter we derive a spectroscopic distance which
we compare to the parallax measurements of {\sl Hipparcos}. As can
seen in Fig.~\ref{fgk_acc} the root mean square scatter is
$3.5\pm3.0$ for the analysis of the eight stars discussed here.

\section{TLS radial velocity monitoring}
The radial velocity (RV) search program of young and active stars of
the TLS described by \citep{hatzes03}. For this program we use the
2-m-Alfred Jensch telescope of the TLS at Tautenburg which is equipped
with an \'echelle spectrograph with resolving power of
$\Delta\lambda/\lambda =67000$.  During the observations an iodine
absorption cell is placed in the optical light path in front of the
spectrographs slit. The resulting iodine absorption spectrum is then
superposed on top of the stellar spectrum providing a stable
wavelength reference against which the stellar RV are measured. In the
first step, the spectra are bias-subtracted, flat-fielded, and
extracted using standard IRAF routines.
        
In the second step the RVs are calculated by modelling the observed
spectra with a high signal-to-noise ratio template of the star
(without iodine) and a scan of our iodine cell taken at very high
resolution with the Fourier Transform Spectrometer of the
McMath-Pierce telescope at Kitt Peak. The latter enables us to compute
the relative velocity shift between stellar and iodine absorption
lines as well as to model the temporal and spatial variations of the
instrumental profile. See \citet{valenti95} and \citet{butler96} for a
description of the principles behind this technique.  RV measurements
have been made at TLS since 2001 and these show that we can achieve a
routine RV precision of $\approx 3$\,m\,s$^{-1}$.

\subsection{HIP~26779, [GKL99] 145, HD 37394}

\begin{figure}
\includegraphics[width=0.35\textwidth, angle=270]{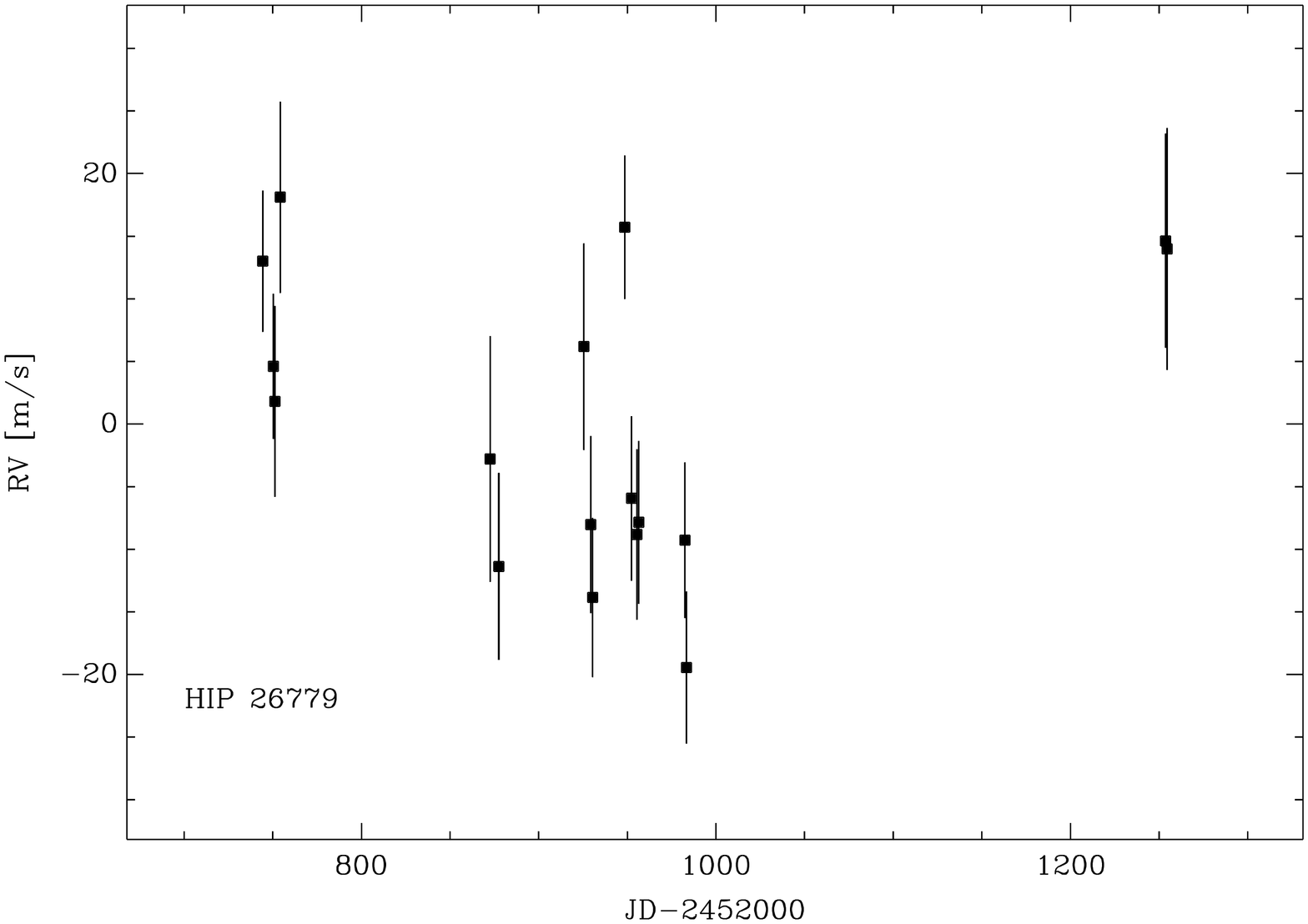}
\caption{RV measurements of HIP~26779 at the TLS. 
  The amplitude of the RV variation are surprisingly small for an
  active star.}
\label{fig:rv_hd37394}
\end{figure}

\citet{Gaidos1998} give a rotation period of $11.02\pm0.05$\,days
derived using the Ca H \& K activity indicator with a constant
RV. \citet{Montes2001} show that the star is a member of the local
association (Pleiades, $\alpha$ Per, M34, $\delta$ Lyr, NGC 2516,
IC2602) and it has an age between 20-150 Myr, thus it is relatively
young and we therefore expect a high level of activity but observe a
rather modest level (see Fig.~\ref{fig:rv_hd37394}): the variance is
only 11.8\,m\,s$^{-1}$ which should be compared to the average error
of the measurement of 7.2\,m\,s$^{-1}$.  We explain this by the
relatively slow rotation which is indicated by the long rotation
period and $v \sin i$. Despite the low amplitude we find a possible
period of about 14 days of the RV-variations.

\subsection{HIP~73555}
\begin{figure}
\label{fig:rv_hip73555}
\includegraphics[width=0.35\textwidth, angle=270]{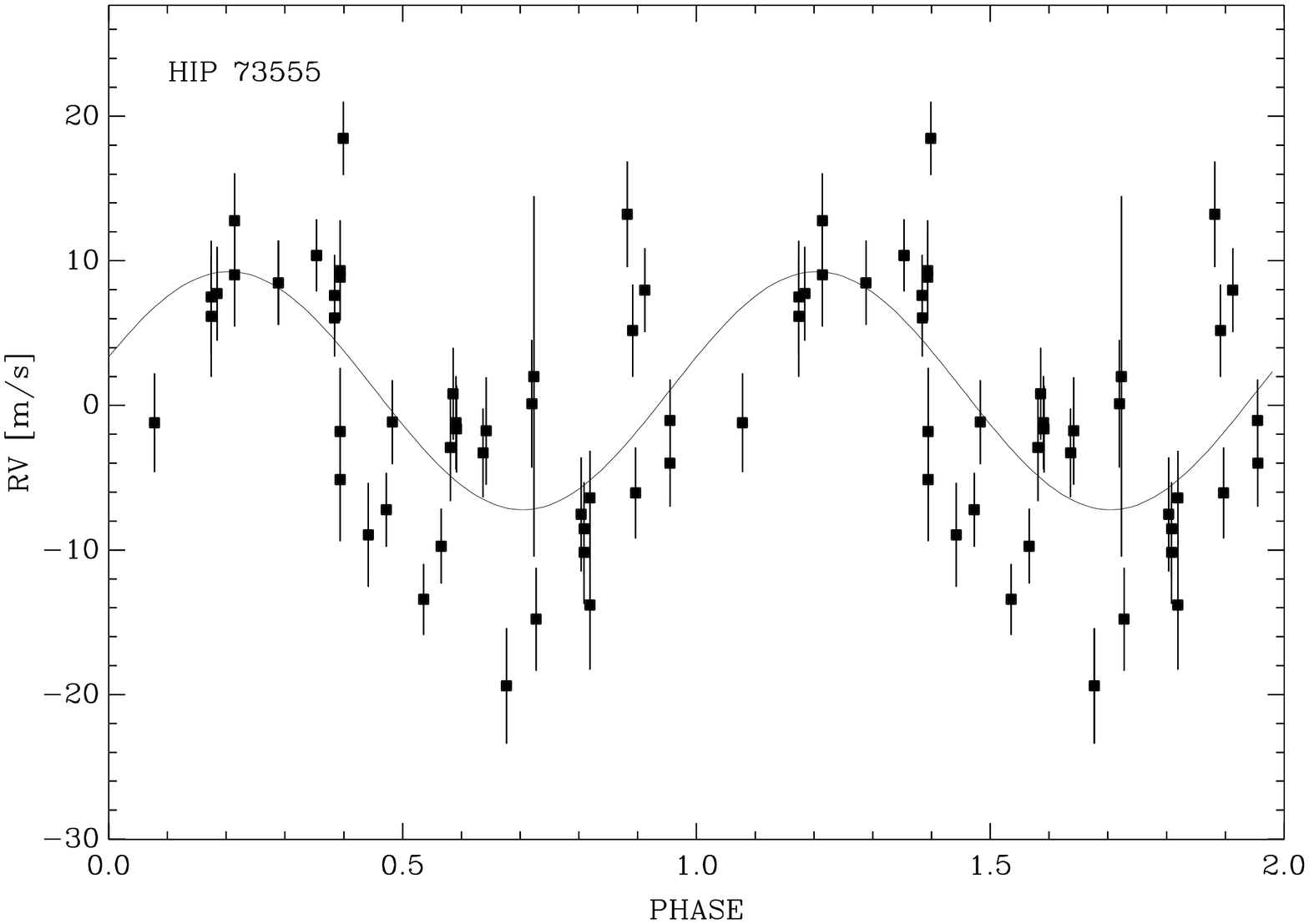}
\caption{RV measurements of HIP~73555 in the years 2001,
  2002, and 2003 at the TLS. The RV variation seen is caused by
  surface features and the period of those variations is that of the
  rotation period of the star.}
\end{figure}

The subgiant HIP~73555 is an interesting testcase for RV
measurements, because as mentioned above there are interferometric
radius measurements available. As mentioned above the radius is
$R=21.5\pm1.0\,{\rm R}_\odot$. From the spectra we determined a $v\sin
i$ of $4.10 \pm 1.0 $\,km\,s$^{-1}$ (Tab.~\ref{tab:parameters}).  We
observed the star for three years, and obtained 42 RV-measurements.
In this data, we find a sinusoidal variation with a period of
$200\pm15$ days and a semi-amplitude of about 10\,m\,s$^{-1}$ (see
Fig.~\ref{fig:rv_hip73555}). The obvious explanation of this variation
is simply the presence of stellar surface features in conjunction with
the rotation of the star. Under this assumption, we calculate an
inclination of $28\pm6^o$

\subsection{HN Peg}
\begin{figure}
\includegraphics[width=0.35\textwidth, angle=270]{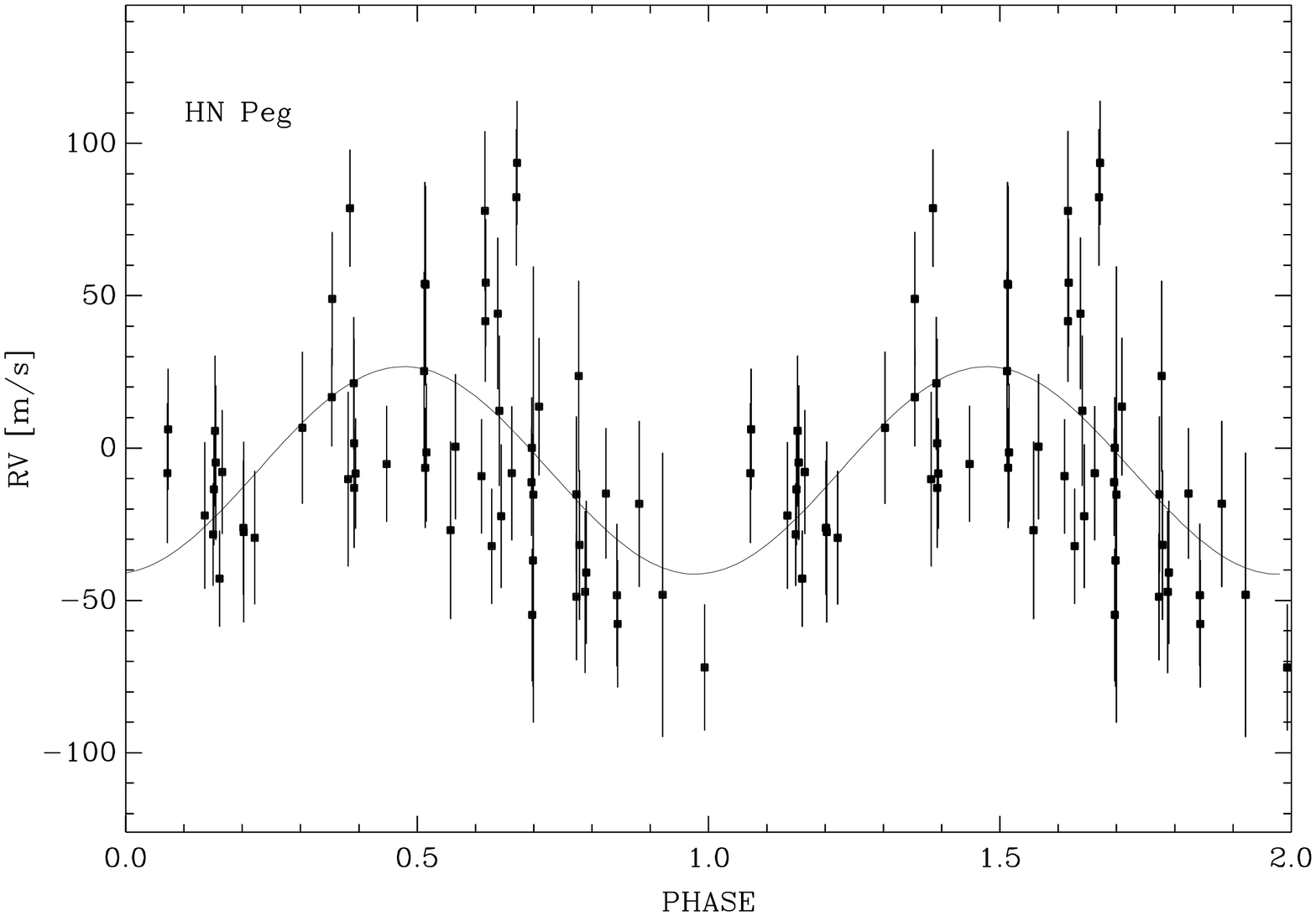}
\caption{RV measurements of HN Peg in the years 2001, 2002, and 2003 
  at the TLS. The RV variation seen are caused by surface features and
  the period of those variations is that of the rotation period of the
  star.}
\label{fig:rv_hnpeg}
\end{figure}

\begin{figure}
\includegraphics[width=0.35\textwidth, angle=270]{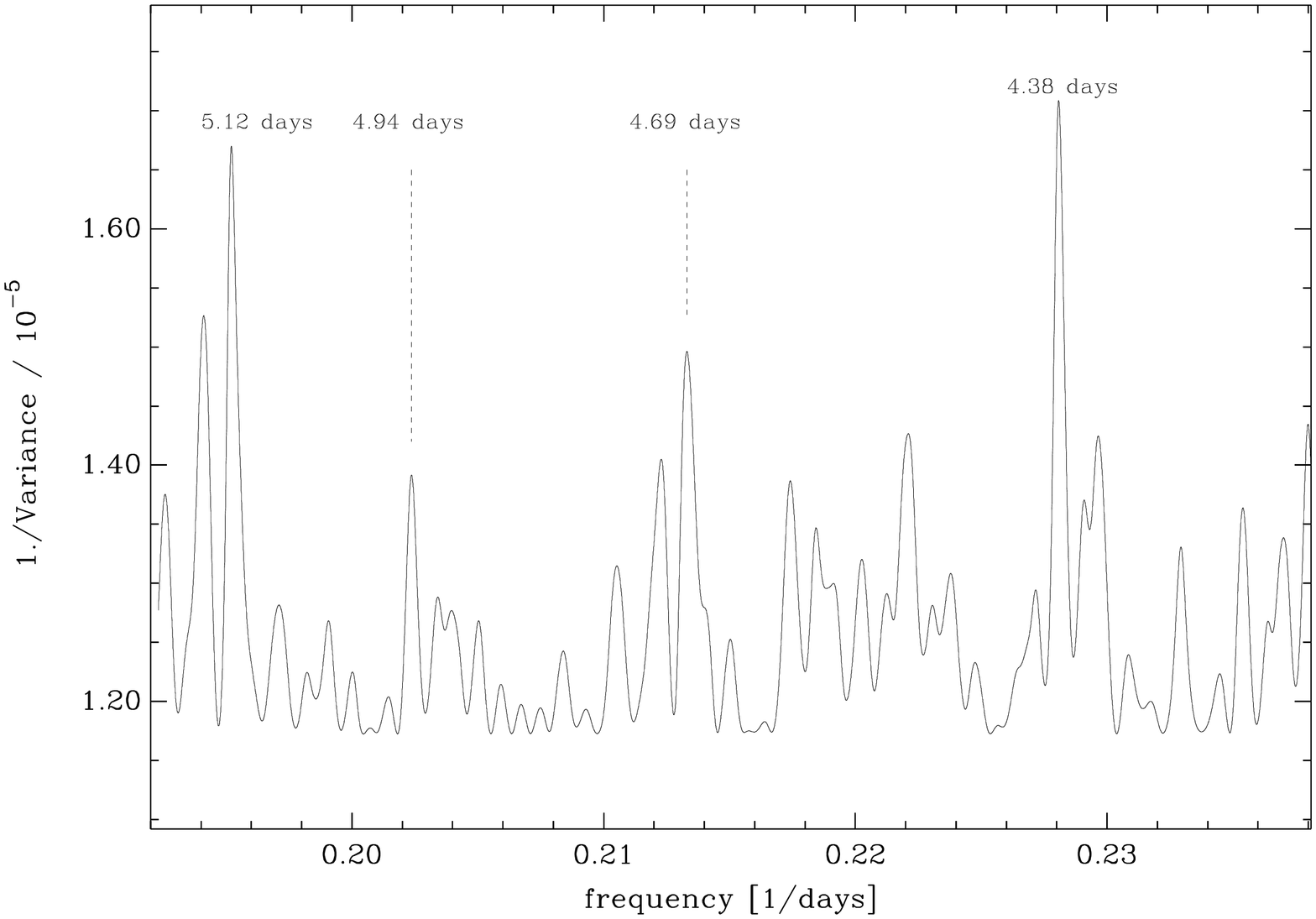}
\caption{Periodigram of the RV measurements of HN Peg. We 
         find several possible periods.}
\label{fig:period_hnpeg}
\end{figure}

In the Lick planet search \citep{Cumming1999} HN~Peg was observed
with a precision of $\sim 5$\,m\,s$^{-1}$, but no significant RV variations
were detected. However, we find sinusoidal variations (see
Fig.~\ref{fig:rv_hnpeg}) with a period of $4.38\pm0.15$\,days and a
semi-amplitude of 30\,m\,s$^{-1}$ which we explain by surface features
which are common for a young star. In addition to this we find less
significant sinusoidal periods: $5.12\pm0.15$\,days,
$4.94\pm0.15$\,days, and $4.69\pm0.15$\,days as shown in the
periodigram in Fig.~\ref{fig:period_hnpeg}.

This strange behaviour agrees well with the results of the photometric
monitoring in the years 1992-1998 of \citet{you2005}. They detect a
period of $5.1348\pm0.0095$\,days which was seen throughout their
entire observing time and additional periods in the range between
4.4 and 4.8 days. They explain this with the activity cycle of the
star in conjunction with differential rotation.

\citet{donahue96} monitored the flux-variations of Ca~H \& K of
HN~Peg for 12 years, and find periods in the range between 4.57
to 5.30 days.  Additionally \citet{frasca2000} measured periodic
H$\alpha$ and Ca~H \& K flux variations, and variations in the
Str\"omgen photometry with a period of 4.74\,days.

Thus, with the RV-measurements we find the same
periods again, as in the photometric data, as well as in the line-flux
variations of the H$\alpha$ and Ca~H \& K.

\subsection{HIP 114385}
\begin{figure}
\includegraphics[width=0.35\textwidth, angle=270]{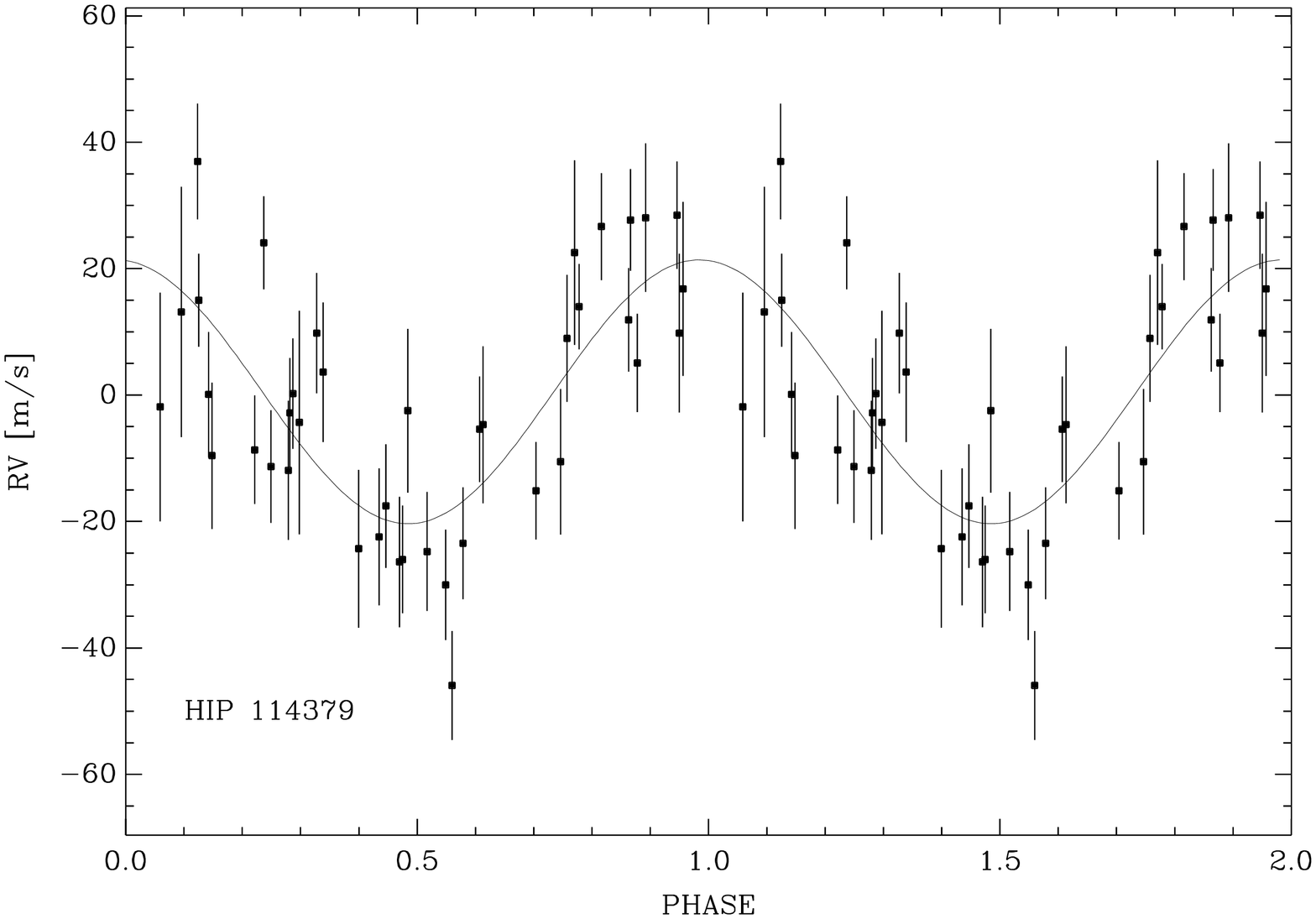}
\caption{RV measurements of HIP~114385 in the years 2001,
  2002, and 2003 at the TLS. The RV variation seen is caused by
  surface features and the period of those variations is that of the
  rotation period of the star.}
\label{fig:rv_hip114379}
\end{figure}

\begin{figure}
\includegraphics[width=0.5\textwidth, angle=0]{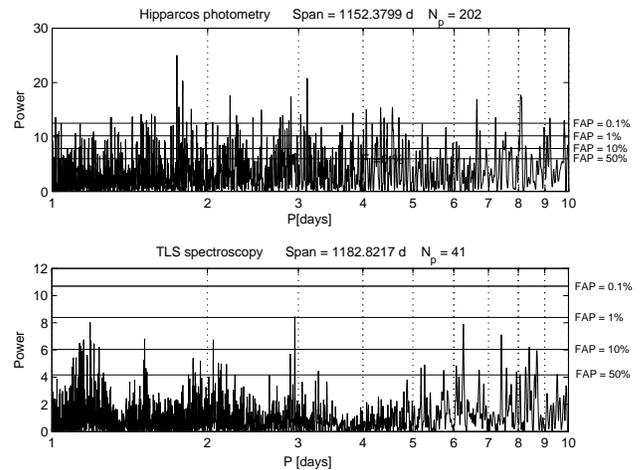}
\caption{Periodigram of the RV-variations and the photometry of
        HIP~114385. Both show that the period is about
        days.}
\label{fig:hip114379_period}
\end{figure}

\begin{figure}
\includegraphics[width=0.5\textwidth, angle=0]{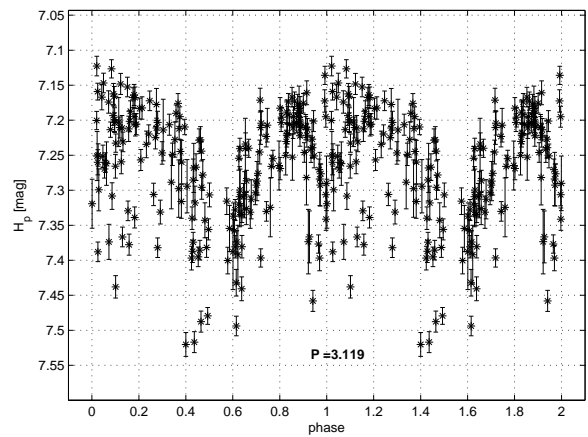}
\caption{Phase-folded  {\sl Hipparcos} light-curve with a period of
  3.119 days of HIP 114385.}
\label{fig:hip114379_phot}
\end{figure}

We monitored the star in 2001-2003 and it showed periodic RV variation
of $2.95\pm0.05$ days and a semi-amplitude of about 20\,m\,s$^{-1}$
(Fig.~\ref{fig:rv_hip114379}). An analysis of the photometric
measurements carried out by {\sl Hipparcos} shows also the same period
of $2.95\pm0.05$\,days.  Fig.~\ref{fig:hip114379_period} shows the
periodigram of the photometric data compared to our RV measurements.
The Fig.~\ref{fig:hip114379_phot} shows the phase-folded
light-curve. We conclude that the RV variations are cause by surface
features.

\section{Discussions}

{\bf HIP~1803} is a member of the Hyades supercluster. All
spectroscopic signatures confirm that the star is ZAMS. For HIP~1803
the newly derived spectral parameters differ quite a lot for older
estimates. Regarding the difference between the spectroscopic distance
and the parallax measurement of {\sl Hipparcos} we are confident that
our estimates are reliable.

{\bf HIP~18512 A} is a member of the Ursa Major
association. The kinematics, the lithium abundance, as well as the
$[Fe/H]$, and the $v \sin i$ support this conclusion.

{\bf HIP~26779} is a young star. The lithium in the atmosphere as well
as its membership to the local association support the young age. But we
are not able to give a definite answer to how old this star really
is. We can exclude a planetary companion with a mass $\geq$ 0.11
M$_{\rm Jupiter}$ with a period $\leq$ 10\,days.

{\bf EK~Dra} is a binary with an orbital period of
$45\pm5$\,yrs. It is one of the young stars where the true masses can
be derived using only Keplers laws. The masses are
$0.9\pm0.1$\,M$_\odot$ and $0.5\pm0.1$\,M$_\odot$ respectively
\citep{koenig2005}. Our spectral synthesis analysis confirms the
mass of the primary. The light contribution of the secondary to the
spectrum of the primary is negligible so that the analysis is not
affected.  We can exclude a planetary companion with a mass $\geq$ 1.0
M$_{\rm Jupiter}$ with a period $\leq$ 10\,days.

{\bf HIP~73555} is a young post-main-sequence giant and not a
pre-main-sequence star. Therefore it should not be listed in the
\citet{gershberg} catalogue. The conclusions for this star is: it is a
young post-main-sequence star with only 240\,Myrs but due to its mass
it has already evolved past the main-sequence and thus does not belong
to this sample. But since the star is relatively young it might anyway
be interesting to search for planets in its vicinity.  For this star,
we can exclude a planet with a mass of more than 0.6 M$_{\rm Jupiter}$
with a period of less than 200 days ($\sim$ one AU).

The other stars are all ZAMS stars. The flare activity is probably
related to the differential rotation and the magnetic field of the
star. Though the spot coverage of {\bf EK~Dra} is about 1/3 of the
visible surface at some phases of rotation and activity, these kind
spots are not know on the surface of the Sun.

On the other flare stars analysed on course of this work we realise
that the H$\alpha$ cores are often filled-in and that the temperature
estimated from H$\alpha$ is if the core is filled by more than 10\%,
20-40\,K lower than the H$\beta$ temperature. In those cases we give
priority to the H$\beta$ temperature. Especially while analysing date
from {\bf EK~Dra} \citep{koenig2005} we experience these changes
dramatically. We have also seen a temperature difference of 120\,K
measured using only H$\alpha$ at spectra with maximal filling-in and
minimal filling-in of the line core. The H$\beta$-line is not that
strongly filled-in so that the temperature difference is within the
errorbars not measurable.

The lithium abundance of {\bf HN~Peg} and {\bf EK~Dra} exceeds the
lithium abundance of other G-type stars, even for their young age. It
is interesting to note that the two stars with the lithium
overabundance are also the most active stars of our sample.  It is
will known that active, late type stars have an overabundance of
lithium \citep{morel2004}.  We thus speculate that in this star
lithium is generated during flare events \citep{Ramaty2000}.  {\bf
HN~Peg} and EK~Dra do not show other peculiarities in its
spectrum. They are otherwise typical ZAMS stars.  We can exclude a
planetary companion with a mass $\geq$ 0.4 M$_{\rm Jupiter}$ with a
period $\leq$ 10\,days around HN~Peg.

{\bf HIP~114385} and {\bf HIP~114376~A\&B} The stars
form a hierarchical triple system. The spectroscopic and AO binary
HIP~114376~A\&B is the star appearing in the flare star
catalogue. Maybe the binarity triggers the activity. We cannot analyse
the binary but the analysis of the companion is possible and we can
then assume a common formation scenario and thus the same age, and
metallicity.

As a general remark we conclude that all these stars are ZAMS stars
making the older than previously suspected by having only the
information about the flaring activity. Due to their proximity to the
sun they are all good targets for direct imaging searches for planets
which are already ongoing. RV monitoring for the planet search are
challenging as the example of EK~Dra shows due to the flaring activity
and stellar spots on the surface.

\section{Conclusion}
The stars all have about solar metallicity. It is interesting to note
the two most active stars of this sample have an overabundance of
lithium. We deduce a mean $v \sin i = 6.0\pm4.6$\,km\,s$^{-1}$ of our
sample stars. In all monitored stars we detect periodic RV
variations. The semi-amplitude is typically 20\,m\,s$^{-1}$. This
corresponds to the activity level seen in the RV amplitude of
$\epsilon$~Eri \citep{hatzes2000}. Around this star it was possible to
detect a planet of $m \sin i = 0.86$\,M$_{\rm Jupiter}$ which caused a
semi-amplitude of 19\,m\,s$^{-1}$, an eccentricity of $0.608\pm0.041$,
a semimajor axis of 3.4\,AU and a period of $2502.1\pm20.1$\,days
\citep{hatzes2000}. For HIP~26779, EK~Dra, HIP~73555, HN~Peg,
HIP~114385 we can exclude 51~Peg like planets: Our upper limits are
0.11, 1.0, 0.6, 0.4 M$_{\rm Jupiter}$, respectively.

\section*{Acknowledgements}
  This research has made use of the SIMBAD database, operated at CDS,
  Strasbourg, France. B.K. wishes to thank Klaus Fuhrmann for the
  discussions. We acknowledge the help of Ana Bedalov and Matthias
  Ammler with some of the observations and the technical support of
  the staff of the Alfred Jensch telescope. The analysis of some of
  the data was carried out as part of B.K.s PhD thesis at the
  Max-Planck-Institut f\"ur extraterrestrische Physik in Garching. We
  wish to thank the referee R.E. Gershberg for his comments to improve
  the work and for further suggestions how to continue this work
  succesfully.

\bsp

\label{lastpage}

\end{document}